\documentstyle[aps,prl,twocolumn,epsf]{revtex}
\input{epsf}
\begin{document}
\title{Correcting  the quantum clock: conditional sojourn times}
\author{S. Anantha Ramakrishna\footnote{e-mail: sar@rri.res.in}
 and N. Kumar\footnote{e-mail: nkumar@rri.res.in}}
\address{Raman Research Institute, C.V. Raman Avenue, Bangalore 560 080, India}
\twocolumn[
\widetext
\begin{@twocolumnfalse}
\maketitle
\begin{abstract}
Can the quantum-mechanical sojourn  time be clocked without the clock affecting
the sojourn  time? Here we re-examine the previously proposed non-unitary
clock,  involving absorption/amplification by an added infinitesimal imaginary
potential($iV_{i}$), and find it {\it not} to preserve, in general, the
positivity of the sojourn time, conditional on eventual reflection or
transmission. The sojourn time is found to be affected by the scattering
concomitant with the mismatch, however small, due to the very clock
potential($iV_{i}$) introduced for the purpose, as also by any prompt
scattering involving partial waves that have not traversed the region of
interest. We propose a formal procedure whereby the sojourn time so
clocked can be corrected for these spurious scattering effects.
The resulting conditional sojourn times are then positive definite for
an arbitrary potential, and have the proper high- and low-energy limits.\\

\noindent PACS Number(s): 03.65.-w, 42.25.Dd, 73.40.-c\\

\end{abstract}
\end{@twocolumnfalse}
]
\narrowtext

\begin{figure}[h]
\epsfxsize=230pt
\begin{center}{\mbox{\epsffile{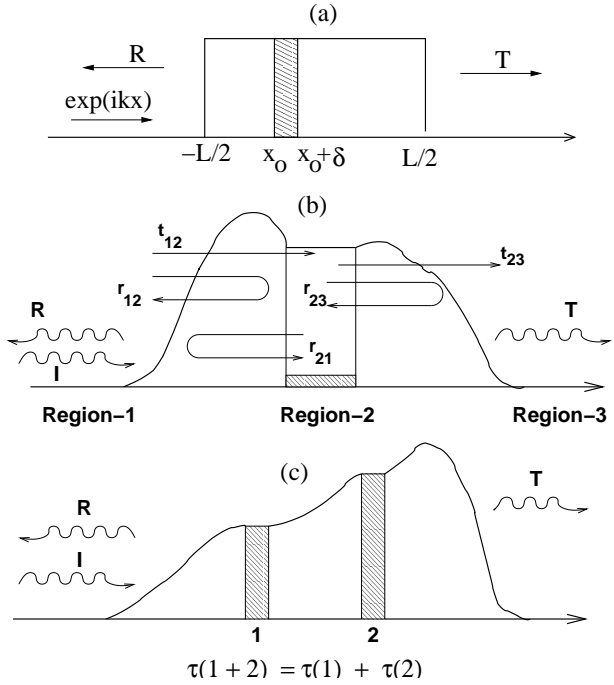}}}
\end{center} \caption{The potentials considered here (a)The rectangular barrier
(b)The region of interest in a region of constant potential is bounded by two
arbitrary potentials whose scattering matrices are shown and (c) Shows two such
regions, the local sojourn times of which add up to  give the total sojourn
time. The hatched region indicates the presence of the clock potential
($iV_{i}$).}
\label{arbitpots} \end{figure}

The time scales associated with the motion of a deformable object, such as a 
quantum-mechanical wave packet, scattered by a potential are 
operationally not context-free and raise some fundamental questions of
interest for mesoscopic systems (for recent reviews see
\cite{landauer94,chiao97,HandS}). 
In the present work, we will be interested in a
physically  relevant time scale that may aptly be called the
(conditional) sojourn time which literally measures the time of sojourn of a
particle in the spatial region of interest, under given conditions of
scattering. Thus, one speaks of the conditional sojourn time --- conditional on
eventual reflection or transmission in the 1-D case. The unconditional sojourn
time, irrespective of eventual reflection/transmission is then the dwell time.
(We could also generalize the sojourn time to include the dwell time for a
particle initially prepared in a spatially confined state --- this covers the
decay time of a metastable state.) Operationally, the sojourn time can be
defined meaningfully by invoking a mathematical artifice  called the ``clock''
involving, e.g. the attachment of an extra degree of freedom that co-evolves
with the sojourning particle. The evolution may be a periodic one as in
the case of the unitary Larmor clock\cite{baz,buttiker83} that involves the
precessional angle accumulated by a spin associated with the particle in an
infinitesimal magnetic field introduced for this purpose. Another
`crossover clock' involves the time-harmonic modulation of the
scattering potential. Here the timescale of traversal is identified with a
certain adiabatic to non-adiabatic crossover phenomenon that occurs when the
traversal time matches the period of modulation\cite{buttiker82}. Both of the
above `clocks' yield a timescale $\tau_{BL} = mL/\hbar \kappa$, called the
B\"{u}ttiker-Landauer time for tunneling through a nearly opaque rectangular
barrier of width $L$, where $\hbar\kappa$ is the magnitude of the imaginary
momentum under the potential barrier. There is yet another `clock', the
e-folding `non-unitary' clock
\cite{golub90,buttiker90,sardelay}, wherein an infinitesimal
imaginary potential is introduced over the spatial interval of interest, and
the conditional sojourn time for reflection/transmission is then calculated as
the derivative of the logarithm of the reflection($\vert R \vert^{2}$) / the
transmission($\vert T \vert^{2}$) coefficient with respect to the imaginary
potential, in the limit of the latter tending to zero, i.e., as $ \tau_{s}^{R}
= \hbar/2~\lim_{V_{i} \rightarrow 0} \partial \ln \vert R \vert^{2} /\partial
V_{i}$ and $ \tau_{s}^{T} = \hbar/2~\lim_{V_{i} \rightarrow 0} \partial \ln
\vert T \vert^{2} /\partial V_{i}$, respectively. Recently, this was used by us
successfully to calculate the probability distribution of the sojourn time for
reflection from a long one-channel random potential \cite{sardelay}, where the
reflection coefficient is unity with probability one. The above clocks, namely
the periodic Larmor clock and the e-folding non-Unitary clock,  however, do not
yield a positive-definite sojourn time in general. Thus for instance, the
conditional sojourn time, for reflection say, so obtained turns out to be
negative for certain deterministic potentials\cite{golub90,HandS}. Two explicit
cases for which we have verified this negativity are (i) the local sojourn time
for reflection in a sub-interval $[x_{0},x_{0}+\delta]$ of a rectangular 
barrier,  as in
in Fig. \ref{arbitpots}(a) (even
though the total sojourn time in the entire potential region, i.e., in the
inteval $(-L/2,L/2)$, remains positive), and (ii) the total sojourn time for
reflection from a $\delta$-dimer potential consisting of two unequal
$\delta$-potentials  separated by a distance $L$.  Basically, under certain
conditions, the absorptive (amplifying) potential can
counter-intuitively increase (decrease) the reflection or the transmission -- a
manifestation of the Borrmann effect well known in the context of X-ray
scattering \cite{borrman}. The dwell time, however, always stays positive.

Now, admittedly, the conditional sojourn time is not an observable in
the strict sense of quantum mechanics inasmuch as there is no self-adjoint
operator corresponding to it with the sojourn time as its eigenvalue 
\cite{allcock}. To the
best of our knowledge, at least, no such operator has been constructed
satisfactorily so far. This, however, does not diminish the importance of
having a timescale that conforms to our classical intuition of the time of
sojourn determined dynamically for the problem at hand. But, for this we need
to have a {\it prescription} for calculating the sojourn time that yields a
{\it reasonable value} so as to be useful as an estimate. Our criteria for the
above are that the sojourn time so calculated should be (i) real and
positive-definite (unlike the phase delay time of Wigner \cite{wigner} which can be of either sign), (ii) additive for non-overlapping spatial intervals, and
(iii) tend to the proper classical limits ({\it i.e.}, at high energies). Thus,
even if not observable as an operator in quantum mechanics, it shall be a
calculable quantity in terms of which one can discuss timescales (fast or
slow) of physical processes of interest in mesoscopic systems. In the
following, we have explicitly constructed precisely such conditional sojourn
times for scattering in 1-D.

We begin by noting a rather subtle problem associated with the `non-unitary' clock,
as also with the Larmor clock, namely that the very clocking mechanism affects
the sojourn time to be clocked, and it does so 
 finitely even as the perturbing clock potential
is taken to be infinitesimally small. We, therefore, re-examine the
non-unitary clock so as to identify the `spurious' perturbing terms
responsible for the unphysical negativity of the conditional sojourn times. We
then propose a formal and rather simple procedure for extracting the correct
sojourn time. We find that the conditional sojourn time, thus corrected, is
indeed positive definite and satisfies all our criteria of reasonableness. Our
main results for the conditional sojourn times in a 1-D spatial interval
($\Delta$), where the scattering potential ($V_{r}$) is taken to be
flat (constant) are
\begin{eqnarray}
\tau_{s}^{T}(E>V_{r}) &=& \hbar \Delta /2~
\lim_{\xi \rightarrow 0} \partial \ln \vert T(V_{i}=0, \xi) \vert^{2}/ \partial
\xi ,\\
\tau_{s}^{T}(E<V_{r}) &=& i\hbar \Delta /2 ~\lim_{\xi \rightarrow 0}
\frac{\partial}{\partial \xi} \ln
[\frac{T(V_{i}=0,\xi)}{T^{*}(V_{i}=0,\xi)}],\\
\tau_{s}^{R}(E>V_{r}) &=& \tau_{s}^{T}(E>V_{r}) + \tau_{SC}, \\
\tau_{s}^{R}(E<V_{r}) &=& \tau_{s}^{T}(E<V_{r}) + \tau_{BL} .
\end{eqnarray}
Here $E$ is the energy of the wave, $\xi = V_{i}\Delta$ and $\tau_{SC}$
is the semi-classical time in the high energy, {\it i.e.}, $\tau_{SC} =
m\Delta/\hbar k$. The above times are found to be positive, and also additive,
and, therefore, can be used to derive the conditional sojourn times for an
arbitrary potential inasmuch as the latter can be approximated as piece-wise
constant.

In the following, we will consider the case of transmission for the
above-the-barrier wave energy (non-tunneling) as also for the 
sub-barrier wave energy
(tunneling) separately. The case of reflection, where a further
logical refinement of our procedure is required, will be considered
separately later. \\ {\it The case of wave propagation, $E>V_{r}$}:
Let us first consider the case of propagation (non-tunneling).  For this, we
calculate the total transmission and reflection amplitudes from the 
multiple reflections arising from the interfaces of the barrier as shown in
Fig. \ref{arbitpots}(b). In the case of propagation we obtain
 \cite{bornwolf}
\begin{eqnarray}
T &=& t_{12}t_{23} e^{ik'L} + t_{12}r_{23}r_{21}t_{23} e^{3ik'L}\nonumber \\
&+& t_{12}r_{23}r_{21}r_{23}r_{21}t_{23} e^{5ik'L} + \cdots ~~~,\\
\label{partialwaves}
R &=& r_{12} + t_{12}r_{23}t_{21}e^{2ik'L}
+ t_{12}r_{23}r_{21}r_{23}t_{21}e^{4ik'L}
+ \cdots~~~,
\end{eqnarray}
where  $k' = \sqrt{2m(E-V_{r} -iV_{i})}/\hbar$ and, $r_{12}$, $r_{23}$, $r_{21}$
and $t_{12}$, $t_{23}$, $t_{21}$ are
the reflection and the transmission amplitudes at the respective interfaces
(See Fig. \ref{arbitpots}(b)).
The transmission coefficient has a generic form
$ T = \sum_{k} A_{k} e^{i\phi_{k}} e^{\alpha_{k}L}$, where
$A_{k}$, $\phi_{k}$ and $\alpha_{k}$ are real numbers
representing the amplitude, phase and the growth of the partial
waves. Consider now the conditional sojourn time given by the non-unitary clock
as
 \begin{eqnarray}
 \label{tauexpand}
\tau_{s}^{T} &=& \lim_{V_{i} \rightarrow 0} \frac{\hbar}{2E} \frac{1}
{\vert T \vert^{2}} \frac{\partial}{\partial V_{i}} \Big[ \sum_{k}
A_{k}^{2} e^{2\alpha_{k} L} \nonumber \\
&+& \sum_{k \neq l} A_{k} A_{l}
e^{i(\phi_{k}-\phi_{l})} e^{(\alpha_{k}+\alpha_{l})L} \Big].
\end{eqnarray}
The imaginary part $iV_{i}$ of the clock potential modifies the
reflection/transmission coefficients ($r_{jk},t_{jk}$) at the interfaces,
where there is mismatch due to the imaginary clock potential. Now, the
derivative with respect to the imaginary potential would cause terms of first
order in $V_{i}$ to contribute to $\tau_{s}^{T,R}$, even in the limit of an
infinitesimal potential $V_{i} \rightarrow 0$. Thus, the clock modifies
`spuriously' the propagation of the wave itself in a non-trivial manner, in
addition to causing the amplification or attenuation of the wave for which
it was introduced.

This analysis immediately suggests the key to correcting the quantum
clock for the `spurious' scattering. The whole point is that the presence
of the imaginary potential modifies the reflection and 
the transmission coefficients at
any point where the imaginary potential changes abruptly, even if
infinitesimally. We have to, therefore, devise a method by which the clock
potential ($iV_{i}$) causes only the intended effect (amplification/absorption)
without causing the `spurious' scattering, i.e., it must be well apodized. A
 little thought of the perturbative structure of the scattering
processes should convince one that the clock related growth/attenuation
would only involve the paired combination $V_{i}\Delta$ ($\Delta$ being
the spatial interval of interest) while the `spurious' scattering
would involve unpaired $V_{i}$. This motivates the following
formal procedure to eliminate the `spurious' effects. Treating
$V_{i}$ and $V_{i}\Delta \equiv \xi$ formally as independent variables,
we keep $\xi$ constant and let $V_{i} \rightarrow 0$ in the
expression for $T$. The transmission sojourn time is then obtained as
\begin{equation}
\tau_{s}^{T} = \hbar \Delta /2~ \lim_{\xi \rightarrow 0}
\partial \ln \vert T(V_{i}=0, \xi) \vert^{2}/ \partial \xi .
\end{equation}
The same result is obtained also by considering the transfer matrices that
explicitly suppress the `spurious' scattering due to the clock potential
$iV_{i}$.

\begin{figure}[h]
\epsfxsize=200pt 
\begin{center}{\mbox{\epsffile{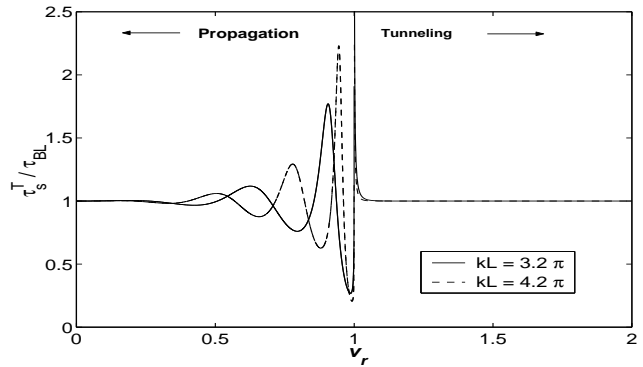}}}
\end{center} 
\vspace{-3mm}
\caption{ The corrected sojourn times for transmission versus  (a)
$v_{r} = V_{r}/E$ for the rectangular barrier
The times are
normalized with respect to the B\"{u}ttiker-Landauer traversal times ($
\tau_{BL}$). \label{sojtimerectdelta}}
\end{figure}

Using either of the procedures, the local transmission sojourn time for the
rectangular barrier region in Fig. \ref{arbitpots}(b) can now be calculated.
Thus, for the case of propagation ($v_{r} <1 $), we have \begin{equation}
\label{tau_T_s} \frac{\tau_{s}^{T}}{\tau_{BL}} = \frac{ (1 - \vert r_{21}r_{23}
\vert^{2})} {1 + \vert r_{21}r_{23}\vert^{2} - 2 \Re
(r_{21}r_{23}e^{2ik_{r}L})}, \end{equation}
where $\Re$ is the real part,
$k_{r} =  \sqrt{2m(E-V_{r})}/\hbar$, and the $r_{jk}$ and $t_{jk}$ are
the scattering amplitudes as before but with $V_{i} = 0$.  We note that since
$\vert r_{jk} \vert < 1$ for any real potential, the above
sojourn time for transmission is always positive.
For the case of the symmetric rectangular barrier
[$r_{21} = r_{23} = (k-k_{r})/(k+k_{r})$],  we explicitly obtain
\begin{equation}
\frac{\tau_{s}^{T}}{\tau_{BL}} = \frac{2(2-v_{r})p}{4 -4 v_{r} +
v_{r}^{2} \sin^{2}(pkL)} ,
\end{equation}
 where $v_{r} = V_{r}/E$ and $p=\sqrt{1-v_{r}}$. In
Fig.~\ref{sojtimerectdelta}, we have shown plots of the transmission sojourn
time for a  rectangular symmetric barrier as a function of the
strength of the scattering potential (for $v_{r}<1$).

We note that the expression given by equation (\ref{tau_T_s}) holds
for a general class of 1-D problems. This is because the $r_{jk}$ can be the
scattering matrix  for any arbitrary potential, with the only condition that
the real potential within the sub-interval, where we seek the time of sojourn,
should be constant (see Fig.~\ref{arbitpots}(b)). This is, however, not a real
restriction as it can be straight-forwardly verified that the local sojourn
times for traversal in different parts of the potential add up to give the
total sojourn time (a schematic is shown in Fig. \ref{arbitpots}(c)).
Since any arbitrary potential
can be constructed out of piece-wise constant potentials (in the limit of the
width going to zero), we realize that the sojourn time for transmission given by
this procedure is positive definite for any arbitrary potential.

{\it The case of wave tunneling, $E< V_{r}$}:
For the case of tunneling, we note that the wavevector becomes imaginary
under the barrier. The real part of the potential sets its own length scale
for the exponential decay / growth with distance under the barrier.
Essentially, the roles of the real part and the imaginary part of the potential
get interchanged. The imaginary part, to first order in $V_{i}$, causes an
oscillation of the wave function with distance. Thus, the paired combination
$\xi = V_{i} \Delta$, will now affect the phase of the wave, rather than its
amplitude. Proceeding as before, we now, get the corrected
sojourn time for transmission in the case of tunneling as the derivative of the
phase with respect to the paired combination $\xi = V_{i} \Delta$:
\begin{equation}
\tau_{s}^{T}(v_{r} >1) = i\hbar \Delta /2 ~\lim_{\xi \rightarrow 0}
\frac{\partial}{\partial \xi} \ln
[\frac{T(V_{i}=0,\xi)}{T^{*}(V_{i}=0,\xi)}] .
\end{equation}
For the general case, as shown in Fig. \ref{arbitpots}(b), we obtain the sojourn
time of traversal as
\begin{equation}
\label{tau_s_tunnel}
\frac{\tau_{s}^{T}}{\tau_{BL}} =  \frac{ (1 - \vert r_{21}r_{23}
\vert^{2} e^{-4k_{r}L})} {1 + \vert r_{21}r_{23}\vert^{2} e^{-4k_{r}L}
- 2 \Re (r_{21}r_{23})e^{-2k_{r}L}},
\end{equation}
where $k_{r} = \sqrt{2m(V_{r}-E)}/\hbar$ now. We note that this traversal
time is positive definite for any arbitrary potential. Again, as before,
the local sojourn
times in different parts of the potential add up to give the total sojourn
time. For the case of the rectangular barrier we obtain
\begin{equation}
\frac{\tau_{s}^{T}}{\tau_{BL}} =  \frac{ (1 - e^{-4k_{r}L})}{1 +e^{-4k_{r}L}
-2[1-8(v_{r}-1)/v_{r}^{2}] e^{-2k_{r}L} } .
\end{equation}
The sojourn time is plotted in Fig. \ref{sojtimerectdelta} (for $v_{r}>1$).
For an opaque barrier ($L \gg k_{r}^{-1}$ or $v_{r} \gg 1$), {\it i.e}, in
the low energy limit, the sojourn time in the above expressions tends to
 the B\"{u}ttiker-Landauer traversal time for tunneling ($\tau_{s}^{T}
\rightarrow \tau_{BL}$). Finally, regarding the local sojourn time in any part
of the rectangular barrier, we find that the ratio of the time spent in the
interval $[x_{0},x_{0} +\Delta]$ to the time spent in the entire barrier is
$\Delta/L$, irrespective of the location $x_{0}$, as is also the case for the
$\delta$-dimer potential. We conclude that in these cases the wave spends
equal amounts of time in equal intervals of the barrier region.

{\it The conditional sojourn time for reflection}:
We now consider the sojourn time for reflection for the case of
 over-the-barrier propagation as well as for the sub-barrier tunneling.
If we look at the  partial wave expansions for the transmission and the
reflection amplitudes in  Eqn.(\ref{partialwaves}), we would realize
one essential difference between the transmission and the reflection. All the
partial waves of the transmitted wave sample the region of interest and
correspondingly pick up the paired combination $\xi = V_{i} \Delta$ in the
amplitude (i.e., the magnitude or the phase). In the case of reflection,
however, there is a partial wave amplitude corresponding to the
reflection from the front edge of the potential upto the region of interest
(see Fig. \ref{arbitpots}(b)),
due to the element $r_{12}$ in the multiple wave
expansion that never samples the region of interest where the imaginary clock
potential is introduced. This part corresponds to the {\it prompt} part of the
reflection. Now, it is clear from the above expressions that this partial
wave interferes with the rest of the partial waves, and thus  affects
the sojourn time to be clocked. 
Arguably, if this prompt partial wave never enters the region where the
imaginary potential is applied, then
the weightage corresponding to this partial wave should be eliminated out
of reckoning. This seems reasonable to us at least in the sense of naive  
realism. This can be accomplished by explicitly
removing the term $r_{12}$ in the right hand side of Eqn.(\ref{partialwaves})
in the 1D case. Thus, we obtain the sojourn time for reflection (for $E>V_{r}$ as
well as for $E<V_{r}$) as
\begin{eqnarray}
\tau_{s}^{R}(E>V_{r}) &=& \tau_{s}^{T}(E>V_{r}) + \tau_{SC}, \\
\tau_{s}^{R}(E<V_{r}) &=& \tau_{s}^{T}(E<V_{r}) + \tau_{BL} .
\end{eqnarray}
The reflection time in this interpretation is the sum of the transmission time
and a propagation time across the sub-interval. Consequently it is always
greater than the transmission sojourn time. But now the reflection time is also
positive definite.

In conclusion, we have pointed out that the non-unitary clock involving the
imaginary potential ($iV_{i}$) can lead to a negative conditional sojourn time
for non-random potentials. This negativity can be traced to the spurious
scattering caused by the very clock potential introduced for clocking the
sojourn time through coherent amplification/attenuation. A simple, formal
mathematical procedure has been given for eliminating 
the effects of this spurious
scattering. In the case of reflection, we further need to remove the {\it
prompt} part of the reflection. With these corrections, the conditional sojourn
times are found to be positive definite and additive, in general. We also find
that the thus corrected non-unitary clock yields the transmission sojourn time
with the proper low-energy limit in agreement with the B\"{u}ttiker-Landauer
traversal time. This problem of the clocking mechanism affecting the time to
be clocked is not special to the non-unitary clock alone. It also affects the
Larmor clock and possibly every clock where the perturbation
due to the clock mechanism couples to the Hamiltonian.
Finally, the conditional sojourn time proposed here, 
while not an observable in the sense of
quantum mechanics, is a calculable intermediate quantity (like the 
matrix element for a transition), and is practically useful 
in deciding {\it for or against} certain conditions of
rapidity. Moreover, it is reasonable in that it is real, positive, calculable,
causally related to the region of interest,
and has the correct classical limit.

\end{document}